\documentclass[runningheads]{llncs}
\usepackage[numbers,sort&compress]{natbib}

\usepackage{graphicx}
\usepackage{caption}
\usepackage{subcaption}
\usepackage{multirow}
\usepackage{xcolor}
\usepackage{booktabs}
\usepackage{amsmath}
\usepackage{hyperref}

\newcommand{\dataset}{mFollowIR}

\begin{document}
\date{}
\title{\textbf{mFollowIR: a Multilingual Benchmark for Instruction Following in Retrieval}}
\titlerunning{mFollowIR: Multilingual Instruction Following in Retrieval}
\author{
    \textbf{Orion Weller}$^{\hspace{.1em}
    \hspace{.1em}{\color{blue}\boldsymbol{\iota}}}$
    \quad
    \textbf{Benjamin Chang}$^{\hspace{.1em}\color{blue}\boldsymbol{\iota}}$
    \quad
    \textbf{Eugene Yang}$^{\hspace{.1em}\color{blue}\boldsymbol{\iota}}$
    \quad \\
    \vspace{.2em}
    \quad
    \textbf{Mahsa Yarmohammadi}$^{\hspace{.1em}\color{blue}\boldsymbol{\iota}}$
    \quad
    \textbf{Samuel Barham}$^{\hspace{.1em}\color{blue}\boldsymbol{\iota}}$
    \quad \\
    \textbf{Sean MacAvaney}$^{\hspace{.1em}\color{blue}\boldsymbol{\lambda}}$
    \quad
    \textbf{Arman Cohan}$^{\hspace{.1em}\color{blue}\boldsymbol{\gamma\hspace{.1em}\alpha}}$
    \quad
    \textbf{Luca Soldaini}$^{\hspace{.1em}\color{blue}\boldsymbol{\alpha}}$
    \quad \\
    \textbf{Benjamin Van Durme}$^{\hspace{.1em}\color{blue}\boldsymbol{\iota}}$
    \quad
    \textbf{Dawn Lawrie}$^{\hspace{.1em}\color{blue}\boldsymbol{\iota}}$
    \vspace{.5em}\\
    $^{\color{blue}\iota\hspace{.1em}}$Johns Hopkins University
    \quad
    $^{\color{blue}\alpha\hspace{.1em}}$Allen Institute for AI
     \vspace{.5em}\\
    $^{\color{blue}\lambda\hspace{.1em}}$University of Glasgow
        \quad
    $^{\color{blue}\gamma\hspace{.1em}}$Yale University
    \vspace{.5em}\\
    \texttt{oweller@cs.jhu.edu}
}
\authorrunning{O. Weller et al.}
 
\institute{}
\maketitle              %
\begin{abstract}
Retrieval systems generally focus on web-style queries that are short and underspecified. 
However, advances in language models have facilitated the nascent rise of retrieval models that can understand more complex queries with diverse intents.
However, these efforts have focused exclusively on English; therefore, we do not yet understand how they work across languages.
We introduce mFollowIR, a multilingual benchmark for measuring instruction-following ability in retrieval models.
mFollowIR builds upon the TREC NeuCLIR narratives (or instructions) that span three diverse languages (Russian, Chinese, Persian) giving \textit{both} query and instruction to the retrieval models. 
We make small changes to the narratives and isolate how well retrieval models can follow these nuanced changes.
We present results for both multilingual (XX-XX) and cross-lingual (En-XX) performance. 
We see strong cross-lingual performance with English-based retrievers that trained using instructions, but find a notable drop in performance in the multilingual setting, indicating that more work is needed in developing data for instruction-based multilingual retrievers.\footnote{We release all code and data publicly at \href{https://github.com/orionw/FollowIR}{https://github.com/orionw/FollowIR}}
\keywords{Instruction Following  \and Multilingual Retrieval \and Cross-Lingual Retrieval \and Evaluation \and Reranking.}
\end{abstract}
\section{Introduction}
Neural Information Retrieval (IR) models have shown large improvements through the use of language model (LM) backbones which are trained on massive amounts of text \citep{craswell2020overview,izacard2021unsupervised,wang2022text}. 
Modern LMs are able to solve a diverse set of tasks through their ability to follow user-provided instructions.

In IR, steerability through instructions presents a unique opportunity to adapt retrieval models to unseen definitions of relevance at inference time~\citep{asai2022tart,instructor_models,muennighoff2024generative,weller2024promptriever}.
Through this approach, users can customize the behavior of systems long after they have been trained.
As such, there has been a flurry of interest in measuring this ability in IR, with several benchmarks being proposed that use instructions/prompts instead of the simple user intents exemplified by MS MARCO \cite{chen2024open,weller2024followir,oh2024instructir,zhao2024beyond}. 

Measuring the ability of IR models to follow instructions in languages beyond English is still understudied, with little to no work on the topic. Given the large amount of non-English speakers in the world, it is important for search systems to be able to recognize and follow complex instructions when the documents and/or queries are in non-English languages.

We seek to rectify this by building an evaluation set for measuring instruction following in three languages (Russian, Chinese, and Persian) called mFollowIR. mFollowIR builds on previous work by proposing a reranking task similar to FollowIR \citep{weller2024followir}, but adapting it to the multilingual setting.\footnote{We define multilingual as being able to handle many languages, and thus perform evaluation on each language monolingually.} We build on top of the TREC NeuCLIR 2022 and 2023 tracks \citep{lawrie2023overview,lawrie2024overview}, using their \textit{narratives} (or instructions given to relevance assessors) as instructions for retrieval models also. 
These instructions are representative of real-world, complex relevance instruction, thus representing a valuable test bed for instruction-following retrieval models.
Our key intuition is to carefully edit narrative in a manner that leads to predictable changes in the set of relevant documents; then we evaluate models on their ability to correctly change relevance ranking based on these edits. 
This approach aims to isolate the instruction-following ability, disentangling it from standard IR metrics which can be confounded with the keyword-matching (or paraphase-matching) abilities of standard IR models.

Our results show that most multilingual and cross-lingual IR models fail to correctly change their relevance scores when given a complex instruction. However, for IR models trained on instructions we see more positive results, indicating that even English-based instruction training data provides benefits to multilingual instruction following. 

In summary, our work offers the following contributions:
\begin{itemize}
    \item Construction and annotation of a new benchmark, mFollowIR, for multilingual instruction following including human annotated edits to the narratives and translations from fluent speakers of Chinese, Russian, and Persian.
    \item An analysis of how English-based instruction-training data impacts cross-lingual instruction following in retrieval, showing strong performance from instruction-trained retrievers.
    \item An analysis on multilingual instruction following in retrieval, finding worse results compared to benchmarks that use English (including cross-lingual mFollowIR), but still generally positive trends for instruction-based training.
\end{itemize}

Overall, our results help provide a method for future research to build more capable retrieval models across languages, with new evaluation data and insight into how to build IR models that can follow instructions in any language.

\section{Related Work}
\subsection{Multilingual Retrieval Evaluation}
Most IR evaluations have assumed an environment of English queries and documents~\citep{craswell2020overview,dang2007overview}. 
However, conclusions drawn from English retrieval do not necessarily transfer to cases where the queries and documents are in non-English languages. 
While there are cross-language and multilingual retrieval question-answering datasets~\citep{clark2020tydi, asai2021xor}, the actual information needs behind each query are unknown (or not documented)~\cite{asai2020challenges}, preventing us from studying approaches that directly interact with the need instead of through a textual query~\cite{broder2002taxonomy}.

Traditionally (since TREC-5 in 1996~\citep{vorhees1996overview}), TREC develops informational search topics with clear documentation of their titles, descriptions, and narratives to ground the scope of the search~\cite{soboroff2021overview}. 
Since the assessors need to understand languages in both query and document languages, developing evaluation collections for multilingual ad hoc retrieval with a traditional TREC topic development approach is more challenging than for English-only retrieval~\citep{lawrie2023overview}. 
There are several publicly available CLIR collections with narratives from CLEF~\cite{clef08adhocoverview, clef09adhocoverview} covering several European languages, but they are generally smaller than modern IR collections and developed by pooling older retrieval systems. 
The recent NeuCLIR collections, developed during the TREC 2022 and 2023 NeuCLIR tracks~\cite{lawrie2023overview, lawrie2024overview}, while covering only Chinese, Persian, and Russian, contain several million news articles in the document collection for each language and judgment pools built from modern neural retrieval models.

\subsection{Language Models and Instructions}
It is now standard for language models to be post-trained with instructions -- called \textit{instruction-tuning} -- to better understand and respond to diverse user requests with fine-grained requirements. Early work on this topic focused on a broad and diverse range of instructions for them to follow \citep{weller2020learning,instructGPT,longpre2023flan,sanh2022multitask}. 

These early efforts have blossomed for more recent models like Llama and Mistral \citep{touvron2023llama,jiang2023mistral} which are quite capable at instruction-following and are indeed used for a variety of applications that LMs were previously unable to do. Along with the capability increases, there have also been a wide range of works focused on measuring their instruction-following ability across different task types \citep{zhou2023instruction,chang2024survey} and domains \cite{yuan2024following}

\subsection{Long Queries in Retrieval}
Much previous work has examined the relationship between query length and model performance \cite{Bendersky2008DiscoveringKC,bendersky2011parameterized,Bendersky2009AnalysisOL,gupta2015information}. In general, studies have shown that increasing the query length results in worse performance. Like many of these previous works, our work also uses TREC narratives as real-world examples of longer queries and shows that models generally perform worse on them. However, in contrast, our work explicitly focuses on evaluating instruction-following.  

\subsection{Instructions in Retrieval}
Although retrieval models often start training from modern-day LMs like Llama, they typically are not trained with instructions. Some of the earliest work on the topic of training with instructions prepended a dataset prefix that defined relevance \citep{asai2022tart,instructor_models} which has continued with more recent models \citep{muennighoff2024generative,moreira2024nvretrieverimprovingtextembedding}. However, these relevance definitions are often generic, short, and do not typically permit lexical change from the version used during training. More recent work has attempted to change that by training IR models that can handle instructions that are longer and more specific \citep{weller2024promptriever} or that use in-context learning  \citep{li2024making}.

\begin{table*}[t!]
    \centering
    \caption{An example of a query and narrative (instruction) from NeuCLIR 2022 in English and Persian, with the original instruction shown and the altered instruction shown as a diff. Note that the instruction change adds extra information that will reduce the number of relevant documents.}
    \vspace{0.5em}
    \includegraphics[width=01.0\linewidth,trim=0.0cm 0.0cm 0.0cm 0cm]{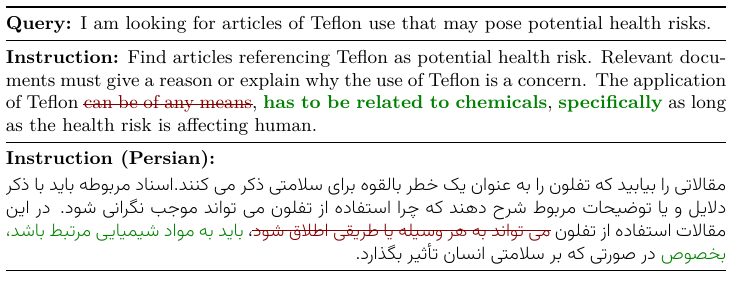}
    \vspace{-0.5em}
    \label{tab:example}
\end{table*}

In evaluation, we see a wide range of new benchmarks being proposed. \citet{oh2024instructir} proposed an MS MARCO \citep{msmarco} based evaluation that appended user personalities to the query (such as \textit{``I am a school teacher looking for..."}) and measured the lowest nDCG score over 10 personalities. \citet{zhao2024beyond,chen2024open} focused on evaluating model's ability to find diverse information, typically from tasks such as argument retrieval. \citet{weller2024followir} proposed to use narratives from TREC tracks to create a method for testing instruction following (FollowIR), measuring a models ability to change based on small edits. Our work uses a similar approach to the one proposed in FollowIR, but approaches it from a multilingual angle using data from the NeuCLIR tracks, as we are not aware of any multilingual evaluation benchmarks for instruction-following.

\section{Dataset Construction}
We construct this dataset following previous work in instruction following in retrieval \citep{weller2024followir} as well as standard IR practices (e.g. pooling) \citep{buckley2007bias,soboroff2003building}. As an overview, we start with the NeuCLIR narratives, alter them, re-annotate the relevant documents, ensure the narrative edit is correctly translated into each language by a native or fluent speaker, and finally pool the top ranked documents (top 1000 docs per query) to create the final reranking task. 

\subsection{Narrative Alterations and Relevant Documents}
\paragraph{Motivation} To test how well models follow the nuances in instructions, we create paired data starting from the NeuCLIR narratives (which contain detailed instructions for relevance, non-relevance and negation \citep{Weller2024NevIRNI}, see Table~\ref{tab:example} for an example) by making a small change to the narrative. However, if we naively changed the narrative, we would have to re-annotate the whole collection -- thus we only add edits that make the instruction more specific (i.e. making some relevant documents non-relevant), ensuring that we only have to re-annotate the \textit{relevant documents}. To use both standard IR metrics (such as nDCG) and instruction-following metrics (p-MRR, see \S\ref{sec:evaluation}), we seek to create an alteration to the narrative such that there are equally as many remaining relevant documents as documents that were previously relevant but have become non-relevant (e.g. for a query with 20 relevant documents, 10 would remain relevant and 10 would become irrelevant). For the example in Table~\ref{tab:example}, the annotator changed ``can be of any means" to ``has to be related to chemicals." This required the annotator to read all the relevant documents beforehand to identify common elements before proposing the edit. We chose to split the relevant documents roughly in half to provide a somewhat equal split of documents to use for different metrics (i.e. nDCG only evaluates relevant documents while p-MRR only evaluates the newly changed non-relevant documents). 

\paragraph{Annotation Procedure} Two annotators with native English proficiency performed this annotation task. Once the annotator altered the narrative, they then read through the list of relevant documents and marked the documents which had newly become non-relevant. Each annotation took approximately 30-60 minutes to iteratively propose a narrative change, read through the relevant documents, and mark them as relevant/non-relevant. We note that one could also look at graded relevance, however, we leave this as future work and leave the document relevance either unchanged or make it completely irrelevant due to the new edit.

\subsection{Ensuring Quality in Translations}
As the alterations were done by English speakers, we needed to be sure that these changes were propagated correctly to the non-English instructions by someone who was fluent in each NeuCLIR language. For each language, we had an annotator who was a native speaker (for Chinese and Persian) or a fluent 2nd language speaker (Russian) translate that alteration from the English narrative into the non-English narrative to ensure high quality non-English instructions.

\begin{table*}[t!]
\centering
\caption{\dataset\ evaluation set statistics. We use a subset of the queries in the TREC NeuCLIR 2022 and 2023 tracks. \textit{$|Q|$} is the character length of the queries and \textit{$|I|$} is the character length of the instructions. Rel. D/Q indicates the number of relevant annotated documents per query in the collection, excluding irrelevant annotations (shown \textit{Before} our annotations and \textit{After} our edits). As designed, there are less relevantly-judged documents in the \dataset\ portion (as the annotations change the relevance of documents on purpose). We show the original (\textit{Orig}) and changed statistics for both the cross-lingual (CL) En-XX and multilingual (Persian, Chinese, Russian) XX-XX settings.
}
\label{tab:statistics}
\vspace{0.5em}
\begin{tabular}{l@{\hspace{0.3cm}}r@{\hspace{0.3cm}}rr@{\hspace{0.3cm}}rr@{\hspace{0.3cm}}rrrr}
\toprule
& & \multicolumn{2}{c}{Rel Docs} & \multicolumn{2}{c}{$|Q|$} & \multicolumn{4}{c}{$|I|$} \\
\cmidrule(lr){3-4} \cmidrule(lr){5-6} \cmidrule(lr){7-10}
Language & \#Q & Before & After & Multi & CL & Orig (en) & CL (en) & Orig (ml) & Multi \\
\midrule
Persian & 40 & 10.8 & 5.4 & 73 & 80 & 397 & 463 & 359 & 415 \\
Chinese & 43 & 10.7 & 5.9 & 24 & 84 & 401 & 456 & 110 & 123 \\
Russian & 40 & 10.0 & 5.6 & 78 & 82 & 371 & 432 & 387 & 458 \\
\midrule
Total & 123 & - & - & - & - & - & - & - & - \\
\bottomrule
\end{tabular}
\end{table*}

\subsection{Pooling Models}
Once we have the edits, we can then pool a variety of search systems to get the top ranked docs (both relevant and non-relevant) for a reranking evaluation. 

\paragraph{Why use a reranking task?}
One of the challenges in evaluating instruction following ability is separating model's ability to perform keyword matching from that of following fine-grained instructions. 
A document that has high lexical (or paraphrase-based) overlap with an instruction is more likely to be relevant than not, although the instruction may provide conditions which exclude the document. Thus, it is the case that most relevant documents to the instruction will be present in the top-K ranked list (where K is large enough); this is confirmed by the empirical finding that most of the relevant documents are already in the model's ranked top-K list -- this is the case in the NeuCLIR 2022/2023 tracks where Recall@1000 is often around 0.9 or greater for the top systems. 

Another reason to use a reranking task\footnote{A reranking task is also naturally quicker to evaluate, which is a nice benefit when using 7B+ parameter models, although not the main reason for choosing it.} is the benefits it brings to a RAG system, where irrelevant documents placed near the top of the ranked list will make it easier for the LM to be distracted and answer incorrectly \citep{shi2023large,yoran2023making,thomas2024large}. Thus, for these long-form and complex instructions, we focus on precision rather than recall, naturally lending itself to a reranking task.

\paragraph{Pooling Procedure} To best evaluate models, the top-K list for reranking should contain hard negatives so that we can effectively discriminate between IR models. Thus we take the top models from the TREC NeuCLIR tracks and pool them, creating a top-K list of 1000 docs/query. We use Naverloo's RetroMAE \citep{lassancenaverloo}, ColBERT-X \citep{nair2022transfer}, CIIR's TransFusion \citep{huangumass}, a PLAID+mT5 pipeline \citep{yang2024hltcoe}, and Naverloo's RankGPT \citep{lassancenaverloo} for the 2023 results pooling and ColBERT-X, UniCamps's P2 reranking system \citep{jeronymo2023neuralmind}, CFDA-CLIP-DQ \citep{ju2022cfda}, KASYS's combined system \citep{abe2022kasys}, and Huawei's system \citep{kamalloo2022huawei} for the 2022 pooling.

\subsection{Overall}
Finally, we combine the 2022 and 2023 queries into one dataset, with one subset per language. Thus, our dataset consists of the original NeuCLIR queries and instructions (in English, Chinese, Russian, and Persian), as well as the paired altered instructions in all four languages. An overview of the dataset statistics can be found in Table~\ref{tab:statistics}. We see that instructions are typically much longer than queries (roughly 5x as long) and that the number of relevant documents per query is roughly 10 before alterations are made, and roughly 5-6 after the alterations to the narratives. Overall, this dataset allows us to examine both cross-lingual and multilingual settings and to compare how models change their ranking based on the targeted alterations to the narratives. 

\begin{figure*}[t]
    \centering
    \includegraphics[width=0.92\linewidth,trim=0.2cm 0.2cm 0.4cm 0cm]{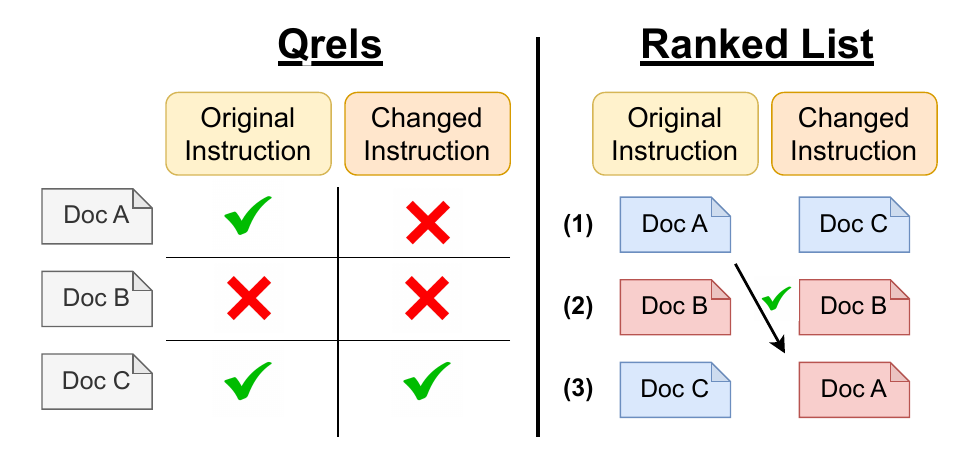}
    \caption{A visual depiction of the pairwise evaluation framework using p-MRR. 
    \textbf{Left}: the original instruction (narrative) is changed to be more specific, making some previously relevant documents newly non-relevant (e.g. \textit{Doc A}). \textbf{Right:} the model is then evaluated on both the original instruction and the changed instruction (along with the query for both). \textcolor{blue}{Relevant docs are in blue}, \textcolor{red}{non-relevant documents are in red}; note that Doc A is relevant for the original instruction but not the changed. p-MRR calculates whether these newly non-relevant documents decreased in rank (in this case, going from Rank 1 to Rank 3 correctly). If the newly non-relevant documents correctly decrease in rank, p-MRR has a positive score (up to 1.0), whereas if the rank increases they have a negative score (down to -1.0), and if there is no change the score is 0 (see \S\ref{sec:evaluation})}
    \label{fig:evaluation}
\end{figure*}

\section{Experimental Settings}
We perform evaluation in two settings (1) Cross-lingually, where the task is English-XX and (2) Multilingually where the task is the same non-English language for both query and documents but is evaluated over several languages. We use the same models and evaluation metrics for both.

\subsection{Evaluation}
\label{sec:evaluation}
We evaluate using nDCG@20 (the default in NeuCLIR, but in our setting we use both query and original instruction as input) and p-MRR (from FollowIR) using p-MRR as our primary metric. The motivation for p-MRR is that models can be strong retrievers based on keywords but fail to consider all aspects of the instruction. As most relevant documents have high lexical/semantic overlap, models can frequently retrieve the relevant documents in the top 20 while still ignoring aspects of the instruction. However, p-MRR explicitly measures their ability to follow instructions by comparing their ranked lists before and after the alteration, sidestepping the issue of the impact of lexical overlap. 

We note that it possible for models to have high nDCG scores but yet low p-MRR scores: this could occur when models ignore the instruction in the input (as it is not needed to find the original relevance annotations) or when models learn to pick out the appropriate keywords from the query and instruction, even if they don't understand the semantic meaning of the instruction. Thus, this motivates our desire to use a metric that can isolate instruction following ability. We note that empirically previous work has found that models use the instruction for keyword matching rather than using them as instructions \citep{weller2024followir}.

p-MRR is calculated by checking the position of the newly non-relevant document (w.r.t. the changed narrative) and comparing whether it went up or down in the new ranked list compared to the ranked list with the original narrative. If the model correctly follows the instructions, the position of the document should decrease (with p-MRR ranging up to 1.0), however, if it ignores the instruction it could stay the same (with a score of 0) or even rank it higher (scores ranging to -1.0, as this is the opposite of what we want). See a visual depiction of this process in Figure~\ref{fig:evaluation}.  More formally, we evaluate p-MRR per query by checking \textit{each} document which is newly non-relevant (using RR as reciprocal rank, $R_{og}$ is the rank when using the original instruction and $R_{new}$ is the new rank):
\begin{equation}
 \text{p-MRR} = 
  \begin{cases}
    \frac{RR_{og}}{RR_{new}} - 1 & \text{if $R_{og} > R_{new}$} \\[10pt]
   1 - \frac{RR_{new}}{RR_{og}} & \text{otherwise} 
  \end{cases}
\end{equation}

For the final p-MRR score, we average first within a given query and then across all queries in the corpora---\textit{i.e.}, macro-averaging across queries, to account for the different number of relevant documents per query. We use PyTREC eval to calculate nDCG@20 \citep{VanGysel2018pytreceval} and use MTEB to calculate p-MRR \citep{muennighoff2022mteb}.

\subsection{Models}
We seek to test a wide variety of models, both bi-encoders and cross-encoders. 

\paragraph{Bi-Encoders} We focus on multilingual models: mContriever \citep{izacard2021unsupervised}, Multilingual E5 of various sizes \citep{wang2022text}, mDPR\footnote{We use the version \texttt{castorini/mdpr-tied-pft-msmarco-ft-all} fine-tuned on MIRACL \citep{101162tacl_a_00595} with tied encoders for ease of use.} \citep{karpukhin2020dense}, and GTE-base-multilingual \citep{zhang2024mgte}.  We also examine the strongest English-only models, which are frequently built on top of multilingual LMs like Mistral: E5-Mistral-Instruct \citep{wang2023improving}, Nomic-Embed \citep{nussbaum2024nomic}, SFR-Embedding-2R \citep{SFR-embedding-2}, GritLM \citep{muennighoff2024generative}, RepLLaMA \citep{ma2023fine}, and Promptriever \citep{weller2024promptriever}.

\paragraph{Cross-Encoders}
We test five cross-encoders: Jina Rerank v2 Multilingual,\footnote{\url{https://huggingface.co/jinaai/jina-reranker-v2-base-multilingual}} BGE Reranker v2 M3 \citep{chen2024bge}, Mistral-7B-Instruct (the LM itself, used as a MonoT5-like reranker) \citep{jiang2023mistral}, mT5-13B \cite{bonifacio2021mmarco}, and FollowIR-7B \citep{weller2024followir}. These models are much more computationally expensive as they compute attention between each query/instruction and document.

\subsection{Hyperparameters and Compute Settings}
We use MTEB for model loading and experiments \cite{muennighoff2022mteb}, using the default parameters for max length. We note that this may cut off some documents for older models with a 512 token context length (such as DPR and Contriever).\footnote{Although this may disadvantage these older models, we note that they are much worse than more recent models, so we include them only as a weak baseline reference.} However, newer models generally have at least 1024 context length or longer, which is long enough for mFollowIR (see Table~\ref{tab:statistics}). We use BFloat16 for the cross-encoder models and for the 7B parameter bi-encoders.

We use one A100 GPU for the evaluations, taking approximately one hour for the smaller models and up to 12 hours for the larger cross-encoder models.

\section{Results}
Overall, we find that models generally struggle at mFollowIR, but that recent work towards instructable English retrievers shows some progress on this task. In each table we show nDCG@20 (when given the query and original instruction), p-MRR, bold the best model per class (i.e., bi-encoder, cross-encoder), and include underlines if the score is statistically similar to the best via a Fisher two-sided randomization test for nDCG@20 and Wilcoxon signed-rank test for p-MRR.\footnote{We use \texttt{ranx} \citep{bassani2022ranx} for the paired Fisher test and \texttt{scipy} \citep{virtanen2020scipy} for the Wilcoxon paired test (which we use due to the potentially non-normal distribution of p-MRR).}

\begin{table*}[t]
\centering
\caption{Results for mFollowIR Cross-Lingual across three language subsets (Persian, Chinese, Russian). Best value in each column is bolded per model type. Models are sorted by average p-MRR, with the highest at the bottom. Underlines signify similarity to the best score in the column via a significance test. We see that models trained on English instructions, such as Promptriever, score highly on p-MRR while other models trained without such instructions do poorly. Note that some models have a high nDCG with a low p-MRR (\S\ref{sec:ndcg_vs_pmrr}).}
\label{tab:mfollowir_cross_lingual}
\vspace{0.5em}
\resizebox{\textwidth}{!}{%
\begin{tabular}{ll|cc@{\hspace{0.2em}}c|cc@{\hspace{0.2em}}c|cc@{\hspace{0.2em}}c|c@{\hspace{0.2em}}c}
\toprule
& & \multicolumn{3}{c|}{Persian} & \multicolumn{3}{c|}{Chinese} & \multicolumn{3}{c|}{Russian} & \multicolumn{2}{c}{Average} \\
\cmidrule(l){3-5} \cmidrule(l){6-8} \cmidrule(l){9-11} \cmidrule(l){12-13}
& Model & nDCG@20 & & p-MRR & nDCG@20 & & p-MRR & nDCG@20 & & p-MRR & nDCG@20 & p-MRR \\
\midrule
\multirow{5}{*}{\rotatebox[origin=c]{90}{\parbox[c]{1.8cm}{\centering \scriptsize Cross-encoder}}}\hspace{0.2em} & Jina-reranker-v2-multi & \textbf{0.505} & & -4.5 & \underline{0.472} & & -1.8 & \underline{0.532} & & -1.9 & \underline{0.503} & -2.7 \\
& BGE-reranker-v2-m3 & 0.344 & & 1.8 & 0.336 & & 2.2 & 0.260 & & 5.4 & 0.313 & 3.1 \\
& mT5-13B-mmarco-100k & \underline{0.492} & & 0.6 & \textbf{0.504} & & 2.8 & \textbf{0.536} & & 6.4 & \textbf{0.511} & 3.3 \\
& Mistral-7B-Instruct & 0.193 & & \textbf{5.8} & 0.417 & & 6.3 & 0.374 & & 4.1 & 0.328 & \underline{5.4} \\
& FollowIR-7B & 0.160 & & -1.1 & 0.327 & & \textbf{11.8} & 0.374 & & \textbf{12.2} & 0.287 & \textbf{7.6} \\
\midrule
\multirow{13}{*}{\rotatebox[origin=c]{90}{\parbox[c]{1.8cm}{\centering \scriptsize Bi-encoder}}} & mContriever-msmarco & 0.129 & & -4.1 & 0.288 & & -0.0 & 0.280 & & -7.5 & 0.232 & -3.9 \\
& mE5-large & 0.307 & & -3.3 & 0.315 & & 0.1 & 0.360 & & -7.0 & 0.327 & -3.4 \\
& mDPR-tied-PFT & 0.069 & & -4.8 & 0.150 & & 1.2 & 0.101 & & -5.7 & 0.107 & -3.1 \\
& mE5-small & 0.230 & & -5.0 & 0.194 & & 0.7 & 0.239 & & -3.2 & 0.221 & -2.5 \\
& SFR-Embedding-2-R & 0.404 & & -7.0 & \underline{0.480} & & 0.6 & \underline{0.530} & & -0.7 & 0.471 & -2.4 \\
& mE5-base & 0.289 & & -3.9 & 0.316 & & 3.4 & 0.307 & & -2.1 & 0.304 & -0.9 \\
& RepLLaMA & 0.263 & & -1.9 & 0.398 & & 1.7 & 0.424 & & -2.3 & 0.362 & -0.8 \\
& GTE-multilingual-base & 0.446 & & -3.7 & 0.414 & & 1.4 & 0.375 & & 0.6 & 0.412 & -0.6 \\
& Nomic-Embed & 0.072 & & -0.5 & 0.129 & & -0.1 & 0.051 & & -0.3 & 0.084 & -0.3 \\
& E5-Mistral-Instruct-7B & \underline{0.450} & & -2.3 & \underline{0.481} & & 0.9 & \underline{0.543} & & 4.4 & \underline{0.491} & 1.0 \\
& GritLM-7B & 0.409 & & -3.5 & \underline{0.494} & & 1.2 & \textbf{0.543} & & 7.1 & \underline{0.482} & 1.6 \\
& Promptriever-Llama2 & 0.281 & & 3.8 & 0.437 & & \underline{7.9} & \underline{0.521} & & \underline{11.4} & 0.413 & \underline{7.7} \\
& Promptriever-Llama3.1 & \textbf{0.522} & & \textbf{8.6} & \textbf{0.504} & & \textbf{9.0} & \underline{0.538} & & \textbf{13.6} & \textbf{0.521} & \textbf{10.4} \\
\bottomrule
\end{tabular}
}
\end{table*}

\subsection{Cross-Lingual Evaluation}
In this setting we test whether models can take English queries and instructions and find relevant non-English documents. We show the results in Table~\ref{tab:mfollowir_cross_lingual} where we see that models generally struggle at this instruction-based task (with scores generally below zero) except for models trained on instructions.

\paragraph{Bi-Encoders} Bi-encoder performance ranges dramatically, with average nDCG@20 scores from 0.08 to 0.52. However, most models perform poorly at following instructions with an average p-MRR of less than zero. The notable exceptions are GritLM, with an average p-MRR of 1.6, Promptriever-Llama2 with 7.7 and Promptriever-Llama3.1-Instruct with 10.4 -- all larger 7B models which have seen instructions (of some kind) during training. 

Similar to the results of the FollowIR paper, we find that some models have high nDCG scores and low p-MRR scores (such as GTE-multilingual-base with an average nDCG of 0.412 and a p-MRR of -0.6) indicating that models use the instruction for lexical overlap rather than as a set of instructions to follow (see \S\ref{sec:ndcg_vs_pmrr} for more discussion on this topic).

\paragraph{Cross-Encoders} We see that the cross-encoders have much strong instruction-following performance in general, enabled through their base model's instruction training and their attention between instruction and documents. We see that Mistral and FollowIR have the highest scores in instruction-following due to their training data (5.4 and 7.6), but have significantly worse performance on standard metrics compared to the highly tuned models from Jina and BGE. Notably, the benefits from FollowIR-7B's small English training data over Mistral are minor, if any, in the cross-lingual setting.

\begin{table*}[t]
\centering
\caption{Results for mFollowIR multilingual across three language subsets (Persian, Chinese, Russian). Best value in each column is bolded per model type. Models are sorted by average p-MRR, with the highest at the bottom. Underline signifies similarity to the best score in the section via a significance test. Scores are worse than in the cross-lingual setting, indicating their English centric bias.}
\label{tab:mfollowir_normal}
\vspace{0.5em}
\resizebox{\textwidth}{!}{%
\begin{tabular}{ll|cc@{\hspace{0.2em}}c|cc@{\hspace{0.2em}}c|cc@{\hspace{0.2em}}c|c@{\hspace{0.2em}}c}
\toprule
& & \multicolumn{3}{c|}{Persian} & \multicolumn{3}{c|}{Chinese} & \multicolumn{3}{c|}{Russian} & \multicolumn{2}{c}{Average} \\
\cmidrule(l){3-5} \cmidrule(l){6-8} \cmidrule(l){9-11} \cmidrule(l){12-13}
& Model & nDCG@20 & & p-MRR & nDCG@20 & & p-MRR & nDCG@20 & & p-MRR & nDCG@20 & p-MRR \\
\midrule
\multirow{5}{*}{\rotatebox[origin=c]{90}{\parbox[c]{1.8cm}{\centering \scriptsize Cross-encoder}}}\hspace{0.2em} & Jina-reranker-v2-multi & \textbf{0.528} & & -1.6 & 0.351 & & \underline{4.4} & \textbf{0.570} & & -2.2 & \underline{0.483} & 0.2 \\
& Mistral-7B-Instruct & 0.111 & & 0.2 & 0.382 & & 1.8 & 0.381 & & \underline{11.5} & 0.291 & \underline{4.5} \\
& BGE-reranker-v2-m3 & 0.459 & & \underline{5.1} & 0.446 & & \underline{2.3} & 0.402 & & 6.2 & 0.436 & \underline{4.5} \\
& mT5-13B-mmarco-100k & \underline{0.472} & & \textbf{5.3} & \textbf{0.517} & & \underline{4.5} & 0.483 & & \underline{7.8} & \textbf{0.491} & \underline{5.9} \\
& FollowIR-7B & 0.090 & & \underline{4.2} & 0.359 & & \textbf{6.7} & 0.406 & & \textbf{12.0} & 0.285 & \textbf{7.7} \\
\midrule
\multirow{13}{*}{\rotatebox[origin=c]{90}{\parbox[c]{1.8cm}{\centering \scriptsize Bi-encoder}}} & mDPR-tied-PFT & 0.149 & & -6.9 & 0.118 & & \underline{-1.0} & 0.162 & & -6.5 & 0.143 & -4.8 \\
& mE5-large & 0.444 & & -8.9 & \underline{0.486} & & -2.4 & 0.430 & & -2.5 & 0.453 & -4.6 \\
& SFR-Embedding-2-R & 0.427 & & -5.4 & \underline{0.488} & & -2.8 & \underline{0.538} & & -1.9 & \underline{0.484} & -3.4 \\
& mE5-small & 0.423 & & -6.4 & 0.361 & & \underline{2.0} & 0.390 & & -3.0 & 0.391 & -2.5 \\
& mE5-base & \underline{0.493} & & \underline{-4.2} & 0.441 & & \underline{0.3} & 0.417 & & -3.5 & 0.450 & -2.5 \\
& mContriever-msmarco & 0.282 & & \underline{-1.3} & 0.375 & & -1.2 & 0.395 & & -4.0 & 0.351 & -2.2 \\
& RepLLaMA & 0.259 & & \underline{-3.3} & 0.405 & & \underline{1.3} & 0.471 & & -1.3 & 0.378 & -1.1 \\
& E5-Mistral-Instruct-7B & 0.439 & & \underline{-2.6} & \underline{0.472} & & -2.8 & \underline{0.531} & & 2.6 & \underline{0.481} & -0.9 \\
& GTE-multilingual-base & \underline{0.483} & & \underline{-4.0} & 0.423 & & \underline{1.1} & 0.420 & & 1.8 & 0.442 & -0.3 \\
& Nomic-Embed & 0.069 & & \underline{0.7} & 0.155 & & \underline{1.5} & 0.130 & & -2.9 & 0.118 & -0.2 \\
& GritLM-7B & 0.391 & & \underline{-0.0} & \underline{0.523} & & \underline{-0.5} & \underline{0.546} & & 1.4 & \underline{0.487} & 0.3 \\
& Promptriever-Llama2 & 0.342 & & \underline{-2.4} & \underline{0.485} & & \textbf{4.7} & \underline{0.515} & & \textbf{10.9} & 0.448 & \underline{4.4} \\
& Promptriever-Llama3.1 & \textbf{0.545} & & \textbf{3.3} & \textbf{0.529} & & \underline{2.8} & \textbf{0.548} & & \underline{9.5} & \textbf{0.541} & \textbf{5.2} \\
\bottomrule
\end{tabular}
}
\end{table*}

\subsection{Multilingual Setting}
In this setting, we explore whether models can search with queries and instructions given in a non-English language, finding documents in that same language. We see the results in Table~\ref{tab:mfollowir_normal}, where we see similar results to the cross-lingual setting, with slightly worse scores across the board. 

\paragraph{Bi-Encoders} Similar to the cross-lingual setting, models perform poorly at instruction following. Notably, all models perform significantly worse at instruction following compared to the cross-lingual setting, with the highest p-MRR average of 5.2 (half of the cross-lingual best). This is likely due to the model's reliance on English data, whereas in this setting there is no English input. Furthermore, there is significant variance in the results, as can be seen by the number of statistically similar results (in underlines).

\paragraph{Cross-Encoders} Cross-encoders performed strongly compared to bi-encoders on the multilingual data as well, with scores as high as 7.7 p-MRR average and good performance for nearly all models. Again we see a large gap between the nDCG@20 scores of the highly tuned retrieval rerankers (like Jina-Reranker-v2) as opposed to FollowIR-7B (0.483 vs 0.285 nDCG@20 averages respectively).

\subsection{Instruction-Following Ability vs Standard Retrieval Ability}
\label{sec:ndcg_vs_pmrr}
As seen in the previous tables, models which are able to retrieve relevant documents may also be ignoring the instructions. 
We saw that models like GritLM, SFR, and Promptriever score the highest on nDCG but only Promptriever is able to score significantly above zero for p-MRR. 
As previous works have shown for English-instruction benchmarks \citep{weller2024followir,Sun2024MAIRAM}, this is likely due to models focusing on keywords as opposed to the entire meaning of the input. Since these TREC collections were gathered via the use of pooling, all documents that were judged (e.g. in the qrels) were found by keyword-based systems. Thus, models can have high nDCG scores on the query/narrative input while only using keywords.

As an example, consider the query and instruction given in Table~\ref{tab:example}. A strong retrieval model tuned to find keyword or paraphrase based matches could likely find many relevant documents using just the keywords ``Teflon" and ``health risk." However, a keyword centric model could miss that it needs to have an AND condition with chemicals \underline{and} humans. This would become especially difficult for keyword-centric models when negation-based clauses are used. Thus, p-MRR is our primary metric for measuring instruction-following, while also showing the original query and narrative nDCG scores (as models should also still be able to maintain general retrieval ability broadly).

\section{Discussion}
\paragraph{Overall, what approaches work best?}
We see that models that trained with some form of instructions did the best in p-MRR, along with larger models (especially 7B+ parameter models) and cross-encoders which can see both query/instruction and document at the same time. The best approaches trained with a large collection of diverse instructions (e.g. Promptriever) or jointly tuned a retrieval model with instruction-following LM data (i.e. GritLM).

Furthermore, we see that approaches that worked well in one language tended to generalize well to other languages.\footnote{With the exception of the FollowIR model, which performed significantly worse on Farsi, likely due to the Mistral model which it is based on.} However, as our dataset does not include low-resource languages, it is unclear whether this will hold when moving to lower-resource languages. Despite this however, the similar performance across languages bodes well for further techniques in instruction-data creation.

\paragraph{Future Work}
Comparing the cross-lingual scores (where models can rely on their English query-based training) vs the multilingual setting (where they must adapt to new languages for queries) we see a wide gap of around 5 p-MRR.  Although p-MRR scores cannot be strictly compared across datasets, we see similar score ranges for our cross-lingual data and English datasets like FollowIR.

This implies that the main gaps are currently (1) a lack of multilingual instruction-based training data and (2) a large gap in performance between models under 1B parameter size and models larger than 1B.

\section{Conclusion}
Our work introduces the first multilingual instruction following benchmark for retrieval, \dataset. \dataset\ is built on top of the TREC NeuCLIR collections and spans the Persian, Chinese, and Russian languages. We evaluate a wide range of retrieval models on this reranking task, both bi-encoders and cross-encoders, and find that the best performance comes from larger models (such as 7B+ parameter models) and from cross-encoders. However, recent efforts to train English instruction following models shows promise even across languages, but shows a noticeable drop when applied to the multilingual setting. Overall, our results show that instruction-training holds promise but crucially must also consider multilingual instruction following data.
\bibliographystyle{splncs04nat}
\bibliography{bibio}
\end{document}